\documentclass[10pt,twocolumn,prd,aps,nofootinbib]{revtex4-1}
\usepackage[english]{babel}
\usepackage{amsmath}
\usepackage{amssymb}
\usepackage{graphicx}

\begin{document}

\title{Dynamical ambiguities in models with spontaneous Lorentz violation}

\author{Yuri Bonder}
\email{bonder@nucleares.unam.mx}
\affiliation{Instituto de Ciencias Nucleares, Universidad Nacional Aut\'onoma de M\'exico\\
Apartado Postal 70-543, Ciudad de M\'exico, Distrito Federal 04510, M\'exico}
\author{Carlos A. Escobar}
\email{carlos.escobar@correo.nucleares.unam.mx}
\affiliation{Instituto de Ciencias Nucleares, Universidad Nacional Aut\'onoma de M\'exico\\
Apartado Postal 70-543, Ciudad de M\'exico, Distrito Federal 04510, M\'exico}

\begin{abstract}
Spontaneous Lorentz violation is a viable mechanism to look for Planck scale physics. In this work, we study spontaneous Lorentz violation models, in flat spacetime, where a vector field produces such a violation and matter is modeled by a complex scalar field. We show that it is possible to construct a Hamilton density for which the evolution respects the dynamical constraints. However, we also find that the initial data, as required by standard field theory, does not determine the fields evolution in a unique way. In addition, we present some examples where the physical effects of such ambiguities can be recognized. As a consequence, the proposals in which the electromagnetic and gravitational interactions emerge from spontaneous Lorentz violation are challenged.
\end{abstract}

% \pacs{11.15.Ex,11.30.Cp,04.20.Ex}

\maketitle

The search for violations of local Lorentz invariance, one of the basic tenets of relativity, is one of the most popular programs to look for quantum gravity effects. This program ranges from fundamental physics, where analyzing if Lorentz violation can arise from a quantum gravity candidate is an active line of research \cite{Pullin}, to experiments, which, in turn, spread over a large spectrum \cite{DataTables}. Therefore, any attempt to limit the possible ways in which our current theories can be consistently extended to incorporate Lorentz violation is of wide interest.

The Standard Model Extension (SME) is a framework designed to parametrize Lorentz violation \cite{SME1,SME2,SME3}. The SME is conceived as an effective field theory and the Lorentz-violating terms in the SME action are assumed to produce small corrections to standard physics. In addition, such terms are of the form of a current, made of conventional fields, coupled with a field that produces Lorentz violation. Interestingly, within the SME, it has been shown that, in curved spacetimes, explicit Lorentz violation generates mathematical inconsistencies \cite{SME3}. Thus, spontaneous Lorentz violation, that is, the situation where dynamical tensor fields subject to special potentials acquire nontrivial ``vacuum expectation values,'' seems to be the only consistent mechanisms to generate Lorentz violation in the SME. 

It should be stressed that spontaneous Lorentz violation was originally discovered in string field theory \cite{LV strings1,LV strings2,LV strings3,LV strings4,LV strings5,LV strings6}. Moreover, it has been claimed that the electromagnetic \cite{NG1,NG3,NG6,NG7,Nambu2} and gravitational \cite{Cardinal} interactions could arise as the Nambu-Goldstone modes associated with such a spontaneous symmetry breaking. In addition, the potentials that are typically used to generate spontaneous Lorentz violation have been utilized in other areas of physics, including, for instance, cosmology \cite{Moffat}.

Being the only viable scheme for Lorentz violation in the SME context, it is paramount to study the self-consistency of the spontaneous Lorentz violation models. Along these lines, the problems of stability \cite{robs, Carroll} and quantization \cite{Hernaski} have been studied for particular models, and some phase space regions have been shown to be unphysical. However, such obstacles are mostly relevant at the quantum level. In this work we uncover additional issues that are relevant even at the classical level. In particular, we analyze if it is possible to construct a Hamilton density for which the associated evolution respects the constraints, and if, given proper initial data, the evolution is uniquely determined. Our results show that such a Hamilton density exists, but that there are evolution ambiguities.

The existence of a Hamilton density that respects the constraints can be analyzed using Dirac's method \cite{Dirac1,Dirac2}. This method also reveals the degrees of freedom of a model, specifying the initial data that should determine the evolution. On the other hand, a model has a well posed Cauchy problem if the evolution is uniquely determined by smooth initial data satisfying the constraints, and if, under initial data changes, such an evolution varies continuously and respecting the causal spacetime structure. In this regard, it has been proved that, in globally hyperbolic spacetimes, any linear, diagonal, second-order, hyperbolic (LDSH) system of differential equations has a well posed Cauchy problem \cite{HawkingEllis,Wald}. However, for the cases where the corresponding equations are not an LDSH system of differential equations, there is no generic method to determine if a theory has a well-posed Cauchy problem and one has to prove each property independently, or appeal to counterexamples to show that one of these properties fails.

For definitiveness we explore these issues in models where a real vector field $B^\mu$ generates the spontaneous Lorentz violation, and a complex scalar field $\phi$ represents matter. Notice that such a vector field is known as a Bumblebee field in the SME community. Also, we work in flat four-dimensional spacetime \cite{footnote2}. Concretely, we focus on the following Lagrange density
\begin{equation}\label{LagrangianGeneral}
\mathcal{L}=-\frac{1}{4}B_{\mu\nu}B^{\mu\nu}-\frac{1}{2}V+\frac{1}{2}D_\mu\phi D^\mu\phi^*-\frac{m^2}{2}\phi \phi^*,
\end{equation}
where $B_{\mu\nu}=\partial_\mu B_\nu -\partial_\nu B_\mu$, $V$ is a nontrivial potential, which, at this point, may be regarded as an arbitrary smooth function of $B_\mu B^\mu$, and $m$ is the scalar field mass. In addition, $D_\mu\phi = \partial_\mu\phi - ie B_\mu \phi$ and $D_\mu\phi^* =\partial_\mu\phi^* + ie B_\mu \phi^*$, where $e$ is a real and dimensionless coupling constant and the star represents complex conjugation. Observe that the term in which the two fields couple can be written as $-B^\mu J_\mu$ with $J_\mu = ie(\phi\partial_\mu \phi^*-\phi^*\partial_\mu \phi )/2$ a real current. Moreover, in contrast with the $V=0$ case, the model under consideration has no gauge symmetry, which constitutes a drastic departure from electrodynamics coupled with a complex scalar field.

The equations of motion are
\begin{subequations}\label{eq evolucion generales}
\begin{eqnarray}
 0&=& \partial_\mu \partial^\mu B^\nu- \partial^\nu \partial_\mu B^\mu - V' B^\nu - J^\nu + e^2 |\phi|^2 B^\nu,\ \label{eom B}\\
 0&=& \partial_\mu \partial^\mu \phi - 2ie B^\mu \partial_\mu \phi - e^2 \phi B^\mu B_\mu,\label{eom phi}
\end{eqnarray}
\end{subequations}
where the prime denotes a derivative with respect to $B^\mu B_\mu$. Note that the corresponding equation for $\phi^*$ can be obtained by complex conjugation, which will be the case throughout the paper. Also observe that, if the second term on the right-hand-side of Eq.~(\ref{eom B}) vanishes, Eqs.~(\ref{eq evolucion generales}) would form an LDSH system of differential equations. In fact, even when $V=0 $, the equations of motion are not an LDSH system of differential equations. However, in this case the gauge symmetry can be used to put the equations of motion in the desired form (\textit{e.g.}, by imposing the Lorentz gauge $\partial_\mu B^\mu=0$).

We turn to the construction of a Hamilton density for which the evolution respects the constraints. The canonical momenta associated with $B_\mu$ are given by $\pi^i=B^{i0}$ and $\pi^0=0$, while the canonical momentum associated to $\phi$ is given by $p=(\partial_0 \phi^*+ i e B_0\phi^*)/2$. Since there is a primary constraint, $\chi_1=\pi ^{0}$, it is necessary to employ Dirac's method to construct the Hamilton density we are seeking. Notice that the constraints classification for this model is given in Ref.~\onlinecite{robs}, where, however, the issue of the existence of a Hamilton density that respects the constraints is not considered and the analysis is done at the level of the canonical Hamilton density. For the model at hand, the canonical Hamilton density is
\begin{eqnarray} 
\mathcal{H}&=&-\frac{1}{2}\pi^i \pi_i + \frac{1}{4}B_{ij}B^{ij} - B_0\partial_i\pi^i + \frac{1}{2}V \nonumber \\
&& +2 p p^* - \frac{1}{2}\partial_i \phi \partial^i \phi^*+\frac{m^2}{2}\phi \phi^* \nonumber \\
&& + ie B_0 (\phi p - \phi^* p^*)+ J_i B^i - \frac{e^2}{2}\phi \phi^* B^i B_i.\label{HamiltonianC}
\end{eqnarray}
Requiring that the time derivative of $\chi_1$ also vanishes leads to the secondary constraint
\begin{equation}\label{constraint primer caso}
\chi_2= \partial_i\pi^i- B_0 V' + 2 e {\rm Im}(\phi p) ,
\end{equation}
which can be thought of as a modification to Gauss law. In addition, $\chi_1$ and $\chi_2$ are second class constraints, reflecting the lack of gauge symmetry. It is possible to add to the canonical Hamilton density a term proportional to $\chi_1$, and its factor can be chosen so that the time derivative of $\chi_2$ vanishes. Thus, the Dirac algorithm has been exhausted without finding inconsistencies or introducing additional constraints, proving that, for any potential, it is possible to construct a Hamilton density whose evolution respects the constraints. Moreover, this analysis shows that the model has $5$ degrees of freedom: $3$ associated with the vector field (two second class constraints remove one degree of freedom \cite{Dirac1}) and $2$ from the complex scalar field. This, in turn, implies that the evolution should be determined by the initial values of $B_i$, $\pi^i$, $\phi$, and $p$.

To tackle the Cauchy problem we have to specify the potential; we consider $V=\kappa(B_\mu B^\mu - b^2)^2/2$, where $\kappa$ and $b$ are real positive constants, and which is a generalization of the Mexican hat potential that is widely used in spontaneous Lorentz violation. Notice that the potential minimum corresponds to a timelike vector field. Hamilton's equations of motion are
\begin{subequations}\label{eq evolucion primer ejemplo}
\begin{eqnarray}
\dot{B}_i&=&\partial_iB_0-\pi_i,\\
\dot{\pi}^i&=&\partial_j B^{ji}-\kappa(B_\mu B^\mu - b^2)B^i-J^i+e^2 |\phi|^2 B^i,\\
\dot{\phi}&=&2p^*+ie\phi B_0,\\
\dot{p}&=&-\frac{1}{2}\partial^i\partial_i\phi^*-\frac{1}{2}m^2\phi^*-\frac{ie}{2}\left[\partial_i(\phi^* B^i)+(\partial_i\phi^*)B^i\right]\nonumber\\
&&-ie p B_0+\frac{e^2}{2}\phi^* B_i B^i,
\end{eqnarray}
\end{subequations}
where the time derivative is denoted by an overdot. Recall that, in addition to the above equations, the fields have to satisfy the constraints $\chi_1=\chi_2=0$. At this point, the main obstacle to have a unique evolution can be identified: according to standard field theory, $B_0$ should be fixed by the constraints at the initial data hypersurface. However, for the potential at hand, $B_0 V'$ is not linear in $B_0$, which implies that, in general, there are multiple values of $B_0$ that are solutions to $\chi_2=0$, and which are thus consistent with the same initial data. In other words, the reduced phase space, which is constructed by replacing $B_0$ in Eqs.~(\ref{eq evolucion primer ejemplo}) by a solution of $\chi_2=0$, is not uniquely determined, and standard field theory does not provide a criteria to choose a particular $B_0$ out of the multiple solutions. Moreover, in contrast with electrodynamics, there is no gauge symmetry that could render such ambiguities unphysical.

\begin{figure}[ht]
\begin{center}
\includegraphics[width=\columnwidth]{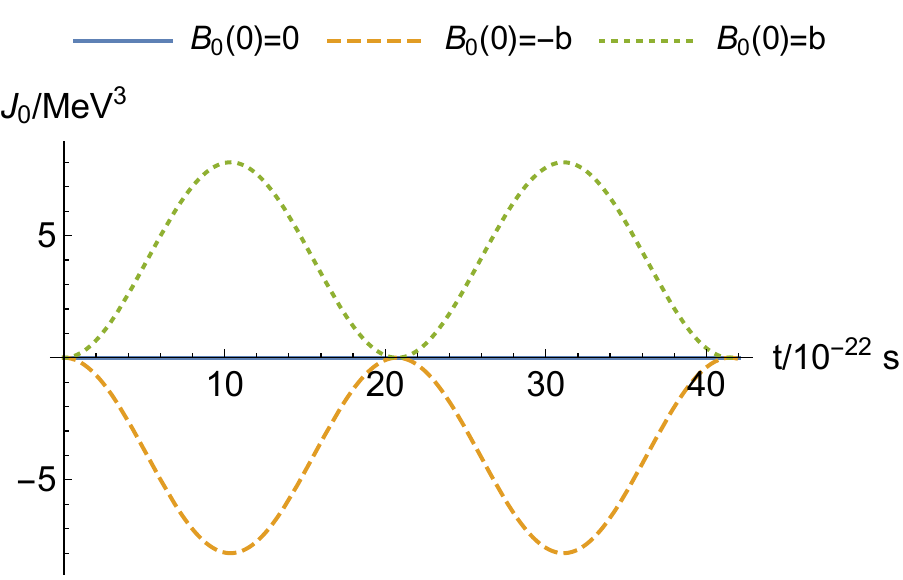}
\end{center}
\caption{\label{Figure}Charge density $J_0$ for the initial data $B_i=0$, $\pi^i=0$, $\phi=0$, and $p=(1+i)\ {\rm MeV} $, and where the parameters of the model are chosen as $m/{\rm MeV}=e=\kappa=b/{\rm MeV} =1$. The solid (blue), the dashed (yellow), and dotted (green) lines represent the situation where $B_0(t=0)$ are, respectively, $0$, $-b$ and $b$.}
\end{figure}

To better grasp the consequences of the issue we are uncovering, we consider a simple example where the initial data, say, at $t=0$, is $B_i=0$, $\pi^i=0$, $\phi=0$, and $p=a$, with $a$ being a complex constant. Note that, being independent on the position on the initial data hypersurface, these initial conditions represent a homogeneous situation. Recall that, according to standard field theory, these are all the conditions needed to specify the fields' evolution. As it can be directly observed by inserting these initial data into Eq.~(\ref{constraint primer caso}), to be compatible with the constraint, $B_0$, at $t=0$, must satisfy $\left(B_0^2 - b^2\right)B_0 = 0$. Thus, $B_0$ on the initial data hypersurface, has three possible solutions that are consistent with the same initial data: $B_0=0,-b,b$. Note that the fact that the initial $B_0$ lies at the extrema of $V$ is a consequence of the particular initial data under consideration. To examine a physical effect that depends on the choice of the initial $B_0$, we solved Eqs.~(\ref{eq evolucion primer ejemplo}) numerically and, in Fig.~\ref{Figure}, we plot the charge density $J_0$, as a function of time, for the different initial $B_0$. This charge density clearly depends on the initial $B_0$: it is zero for $B_0(t=0)=0$, and it has negative and positive oscillations for $B_0(t=0)=-b,b$, respectively. We want to point out that the behavior presented in Fig.~\ref{Figure} does not depend on the precise values of the parameters chosen for the evolution, which are listed in the caption of Fig.~\ref{Figure}.

Spontaneous Lorentz violation can also be produced by potentials containing a Lagrange multiplier $\lambda$. Note that, since $\lambda$ is a dynamical field, the Dirac algorithm analysis presented above cannot be applied in this case. Assuming that the potential has the form $\lambda\tilde{V}$, where $\tilde{V}=\tilde{V}(B_\mu B^\mu)$, the constraints are $\chi_1=\pi^0$, $\chi_2=p_\lambda$,
\begin{subequations}\label{constraints segundo ejemplo}
\begin{eqnarray}
\chi_3&=&\partial_i\pi^i-\lambda \tilde{V}' B_0+ 2 e {\rm Im}(\phi p),\\
\chi_4&=& \tilde{V}.
\end{eqnarray}
\end{subequations}
Clearly, $\chi_1$ and $\chi_2$ are primary constraints, while $\chi_3$ and $\chi_4$ are secondary; all four are second class constraints. In addition, it can be shown that Dirac's algorithm can be exhausted without finding inconsistencies or additional constraints. Thus, the degrees of freedom again are $B_i$, $\pi^i$, $\phi$, and $p$, which can be understood by the fact that the additional degree of freedom associated with $\lambda$ cancels with the constraints it introduces. In particular, this implies that the initial values for $\lambda$ and $B_0$ have to be determined by the initial data through the constraints. Concretely, the initial $\lambda$ must be obtained from $\chi_3=0$, while the initial $B_0$ can be found by solving $\chi_4=0$, which is, in general, nonlinear in $B_0$. Therefore, there are multiple solutions for $B_0$ which are compatible with the same initial conditions, and thus, analogous dynamical indeterminacies from those described above will be present. 

For concreteness, we focus on $\tilde{V}= B_\mu B^\mu - b^2$. Note that, if $e=0$, this model is equivalent to the minimal Einstein-Aether theory in flat spacetime, for which the Cauchy problem is well posed \cite{ether}. However, for the present study, the coupling of the vector and matter fields plays a crucial role, and the result of Ref.~\onlinecite{ether} does not apply. The equations of motion for $B_i$, $\phi$, and $p$ have the same form as their counterparts in Eqs.~(\ref{eq evolucion primer ejemplo}), as can be heuristically argued by the fact that they are independent of $\kappa$, and
\begin{eqnarray}
\dot{\pi}^i&=&\partial_j B^{ji}-\lambda B^i-J^i+e^2|\phi|^2 B^i.
\end{eqnarray}

To analyze a particular case, we take the same homogeneous initial data used above. In this case, the constraints, given in Eqs.~(\ref{constraints segundo ejemplo}), imply that, initially, $\lambda =0$ and $B_0=-b,b$. This is consistent with the fact that $B_0 \neq 0$ to have an invertible matrix formed by the Poisson brackets of the second class constraints, which is a requirement of Dirac's method. Again, there are several initial $B_0$ that are compatible with the initial conditions. To explore the physical consequences, we evolved numerically the equations of motion for $B_0(t=0)=- b,b$. It can be shown that, for the particular initial conditions under consideration, the fields have a similar behavior to the corresponding fields studied above and which are subject to the Mexican hat potential. In particular, the current density also has negative (positive) oscillations for a negative (positive) initial $B_0$.

To summarize, we found that the initial data required by standard field theory, in the spontaneous Lorentz violation models under consideration, does not determine the physical evolution in a unique way. As we mentioned above, this issue can be traced to the nonlinear nature of the constraints, which prevents the initial data, as prescribed by Dirac's method, to determine the remaining components of $B_\mu$. Note that it may be possible to overlook this problem when working to linear order in $B_\mu$. Moreover, observe that such an issue is not a generic feature of models with spontaneous symmetry breaking. A particular instance where no problems arise is when the models have no constraints. A simple example of such a model is provided by a complex scalar field with a standard kinetic term and a potential that drives a spontaneous breaking of the $U(1)$ symmetry. In fact, in this model the equations of motion form an LDSH system of differential equations, implying that it has a well-posed Cauchy problem. 

Different strategies could be considered to reconcile these models with standard field theory. First, to give to $B_0$ the status of a fully dynamical entity, like the other components of the vector field. In this manner, the initial $B_0$ has to be specified as an additional initial data (that has to be compatible with the constraints). This is what one does if one naively applies Lagrange's formalism to the model at hand, and it is justified by the fact that, after all, once one makes enough measurements to know $B_0$, the evolution is completely determined. However, the fact that $B_0$ plays different roles in the Lagrange and Hamilton formulations indicates that the quantization of these models could lead to physically inequivalent theories when different quantization schemes are utilized (\textit{e.g.}, canonical \textit{vs.} path integral quantizations). A second possibility is to construct a criteria to choose, from the alternative evolutions allowed by the initial data, the true physical evolution. One option is to use the initial energy. However, it is easy to find initial data where there are degeneracies, suggesting that constructing such a criteria is, in general, highly nontrivial. Third, to abandon the idea that these models are more fundamental than our current theories, disputing the proposals that electrodynamics and gravity could arise from spontaneous Lorenz violation. After all, these interactions emerge only in a fix gauge, and thus, it is not possible to rely on the standard methods, which crucially depend on the gauge symmetries, to show that the Cauchy problem is well posed. In this regard one can take the point of view that these models are phenomenological and that the fundamental theory, from which these models emerge, should provide a prescription to choose the correct evolution.

There are several generalizations to this work, some of which we have explored. We analyzed the situation where Dirac fermions play the role of matter fields, and which are minimally coupled with the same $B_\mu$ field. It turns out that analogous indeterminacies arise for the two type of potentials considered here. Also, one can use more complicated potentials. Nevertheless, it seems that, if such potentials generate nonlinearities in the constraints, the ambiguities we found should still be present. In fact, it is easy to see that the number of alternative evolutions is directly related with the power in which $B_\mu B^\mu$ and $\lambda$ appear in the potential. It should be emphasized that there are particular initial conditions for which not all the compatible $B_0$ are real, or where there are degenerations, and, in those cases, there is an additional criteria to limit the number of alternative evolutions.

Also, it is possible to consider different kinetic terms for $B_\mu$. It is easy to note that, in flat spacetime, if such terms are restricted to two derivatives and two powers of the vector field, the most general situation is a linear combination of $B_{\mu\nu}B^{\mu\nu}$, $(\partial_\mu B^\mu)^2$ and $\partial_\mu B_\nu \partial^\mu B^\nu$. Observe that $\partial_\mu B_\nu \partial^\nu B^\mu$ is linked with $(\partial_\mu B^\mu)^2$ by a divergence. Clearly, it is difficult to repeat our analysis in a generic situation. However, we found that, for a kinetic term that is only given by $(\partial_\mu B^\mu)^2$, and where the vector field has only $1$ degree of freedom, analogous ambiguities to those we are reporting arise. On the other hand, in the case where the kinetic term is $(\partial_\mu B_\nu) \partial^\mu B^\nu$, there are no constraints and, in fact, the equations of motion form an LDSH system. Thus, this model has a well-posed Cauchy problem. However, the only kinetic term that is independent of the metric-compatible derivative operator is $B_{\mu\nu}B^{\mu\nu}$, and thus, in curved spacetimes, the energy-momentum tensor for the alternative kinetic terms will have two partial derivatives acting on the spacetime metric \cite{Seifert}, which may damage the Cauchy problem. Therefore, it seems that investigating the Cauchy problem for theories with spontaneous Lorentz violation in curved spacetimes would be very interesting, and it could shed light into some longstanding puzzles in the SME \cite{tpuzzle}.

Finally, we would like to point out that, even though this work lies in the classical regime, the issues we are finding could also be present at the quantum level. One may be tempted to think that spontaneous symmetry breaking allows one to freely choose a vacuum, making the ambiguities we found irrelevant. This line of reasoning is motivated by the fact that, when a gauge symmetry is spontaneously broken, selecting a vacuum amounts to fixing a gauge \cite{PerezSudarsky}, which has no physical effects. However, for spontaneous symmetry breaking of nongauge symmetries, including spontaneous Lorentz violation, the vacuum choice can have physical implications, and the type of arguments we uncover could still arise.

\begin{acknowledgments}
We acknowledge partial financial support from UNAM-DGAPA-PAPIIT Project No. IA101116. Also, we thank Quentin Bailey, Carlos Hernaski, Tim Koslowski, Alan Kosteleck\'y, Ralf Lehnert, El\'ias Okon, Rob Potting, Marcelo Salgado, Daniel Sudarsky, Luis Urrutia, and David Vergara for their valuable feedback.
\end{acknowledgments}


\begin{thebibliography}{99}
\bibitem{footnote2}
We use a global coordinate system such that the spacetime metric takes the form $\eta_{\mu\nu}={\rm diag}(1,-1,-1,-1)$. Also, we employ the convention where Greek indices are spacetime indices, while Latin indices represent purely spatial indices. Einstein's summation convention is utilized, and the metric and its inverse lower and raise spacetime indices. In addition, we use units where $c=\hbar =1$.
\end{thebibliography}
\end{document}